\documentclass[twocolumn,epjc3]{svjour3mod}  
\journalname{Eur. Phys. J. C}
\usepackage[utf8]{inputenc}
\usepackage{epsfig,amsfonts,amssymb,latexsym,amsmath,bm,color,array,epsfig,graphics,dsfont}
\usepackage{enumitem}
\usepackage{amsmath}
\usepackage{amssymb}
\usepackage{physics}
\usepackage{lastpage}
\usepackage{graphicx}
\usepackage{subfig}
\usepackage{hyperref}
\usepackage{balance,cite}
\usepackage{booktabs}
\hypersetup{colorlinks=true,linkcolor=blue,citecolor=blue,menucolor=black,urlcolor=blue,filecolor=blue}
\newcommand{\beoa}{\begin{overlayarea}{\textwidth}{\textheight}}
\newcommand{\eeoa}{\end{overlayarea}}

\usepackage{booktabs}
\usepackage{slashed}
\usepackage{rotating}
\usepackage{psfrag}
\usepackage{bchart}
 \usepackage{multicol}
 \usepackage{multirow}

\newcommand{\Order}{\mathcal{O}}

\newcommand{\GeV}{\,\text{GeV}}

\newcommand{\nn}{\nonumber\\}

\newcommand{\Fpigg}{F_{\pi \gamma^* \gamma^*}}
\newcommand{\Fpig}{F_{\pi \gamma \gamma^*}}
\newcommand{\FPigg}{F_{P_i \gamma^* \gamma^*}}
\newcommand{\FPig}{F_{P_i \gamma \gamma^*}}

\newcommand{\F}{\mathcal{F}}

\newcommand{\bsp}{\begin{sloppypar}}
\newcommand{\esp}{\end{sloppypar}}
\newcommand{\toright}[1]{\hspace*{\fill}{\footnotesize{#1}}}

\allowdisplaybreaks[1]

\begin{document}

\title{\toright{\textnormal{PSI-PR-21-13, UWThPh 2021-7}}\\[4mm]Short-distance constraints for the longitudinal component\\[2mm] of
  the hadronic light-by-light amplitude: an update}
\author{ 
G.~Colangelo\thanksref{Bern} \and F.~Hagelstein\thanksref{PSI}\and M.~Hoferichter\thanksref{Bern}\and L.~Laub\thanksref{Bern}\and P.~Stoffer\thanksref{Vienna}
}       
\institute{Albert Einstein Center for Fundamental Physics, Institute for Theoretical Physics, University of Bern, Sidlerstrasse 5, 3012 Bern, Switzerland \label{Bern}
\and
Paul Scherrer Institut, 5232 Villigen PSI, Switzerland \label{PSI}
\and
University of Vienna, Faculty of Physics, Boltzmanngasse 5, 1090 Vienna, Austria \label{Vienna}
}
\date{}
\maketitle

\begin{sloppypar}

\begin{abstract}
We reassess the impact of short-distance constraints for the longitudinal
  component of the hadronic light-by-light amplitude on the anomalous magnetic moment of the muon, $a_\mu=(g-2)_\mu/2$, by
  comparing different solutions that have recently appeared in the
  literature. In particular, we analyze the relevance of the exact axial
  anomaly and its impact on $a_\mu$ and conclude that it remains rather limited.
  We show that all recently proposed solutions agree well within
  uncertainties on the numerical estimate of the impact of short-distance
  constraints on $a_\mu$, despite differences in the concrete
  implementation. We also take into account the recently calculated
  perturbative corrections to the massless quark loop to update our
  estimate and outline the path towards future improvements.
\end{abstract}

\section{Introduction}
The recent measurement of $(g-2)_\mu$ by the Fermilab Muon $g-2$
collaboration~\cite{Abi:2021gix,Albahri:2021ixb,Albahri:2021kmg,Albahri:2021mtf}
has made the discrepancy with the Standard-Model
prediction~\cite{Aoyama:2020ynm,Aoyama:2012wk,Aoyama:2019ryr,Czarnecki:2002nt,Gnendiger:2013pva,Davier:2017zfy,Keshavarzi:2018mgv,Colangelo:2018mtw,Hoferichter:2019gzf,Davier:2019can,Keshavarzi:2019abf,Hoid:2020xjs,Kurz:2014wya,Melnikov:2003xd,Masjuan:2017tvw,Colangelo:2017qdm,Colangelo:2017fiz,Hoferichter:2018dmo,Hoferichter:2018kwz,Gerardin:2019vio,Bijnens:2019ghy,Colangelo:2019lpu,Colangelo:2019uex,Blum:2019ugy,Colangelo:2014qya}
more serious, by bringing it from the $3.7\sigma$ to the $4.2 \sigma$ level when
combined with the previous Brookhaven measurement~\cite{bennett:2006fi}.
The concrete perspective of additional reductions of the experimental
uncertainty in the near future---mainly from subsequent Runs at Fermilab~\cite{Grange:2015fou}, but also the future J-PARC experiment~\cite{Abe:2019thb} using a different technique---makes the need of further theoretical
improvements more urgent. As is well known, the two main sources of
theoretical uncertainties are both hadronic.  The largest one is the
hadronic vacuum polarization (HVP) contribution, and the second hadronic
light-by-light (HLbL) scattering. A lot of work has been devoted to reducing the
uncertainty in the latter by separately analyzing each of the different
contributions to the HLbL amplitude. These involve different intermediate
states and their calculation requires a good understanding of the relevant
physics. The present status has been summarized in the White Paper (WP) on
the Standard Model prediction of $(g-2)_\mu$~\cite{Aoyama:2020ynm}, resulting in a phenomenological estimate in agreement with lattice QCD~\cite{Blum:2019ugy,Chao:2021tvp}. The
behavior of the HLbL amplitude for asymptotic values of its arguments is
fixed by QCD and represents an important, global constraint, which has a
significant impact on the estimate of this contribution. A detailed
understanding of which intermediate states play a role in satisfying this
constraint is crucial to estimate its impact and reduce the overall
theoretical uncertainty.

There are two different regimes of asymptotic momenta and correspondingly
two different constraints. For $g-2$ kinematics, with one photon in the
static limit, the HLbL amplitudes only depend on the squared momenta of the
remaining three photons.  The first regime is when two of them are much
larger than the third, whereas all of them being about equally large
defines the second. They represent two limiting cases of a possible
continuum of short-distance constraints (SDCs) and we will refer to them as
SDC1 ($q_{1,2}^2 \gg q_3^2$, $q_{1,2}^2\gg \Lambda^2_\text{QCD}$) and SDC2 ($q_1^2\sim q_2^2\sim q_3^2\gg \Lambda^2_\text{QCD}$).
Melnikov and Vainshtein (MV) were the first to derive
SDC1~\cite{Melnikov:2003xd} and, in particular, to point out that in the
chiral limit and for asymptotic values of $q_{1,2}^2$, the leading
$1/q_3^{2}$ behavior of the longitudinal part receives no corrections neither at large nor at small
$q_3^2$ values: in other words, the $1/q_3^{2}$ dependence is exact across
the whole range. This is a direct consequence of the 
axial anomaly~\cite{Adler:1969gk,Bell:1969ts,Bardeen:1969md,Wess:1971yu,Witten:1983tw}, see Sect.~\ref{sec:OPE}. The transverse part in turn is constrained by the celebrated
non-renormalization theorems~\cite{Vainshtein:2002nv,Knecht:2003xy} for the vector--vector--axial-vector (VVA) correlator. The SDC2 case was also discussed by MV on the basis of the
quark loop, but its derivation has only recently been put on a firm basis
by using the operator product expansion (OPE)~\cite{Bijnens:2019ghy}.
Moreover, both non-perturbative~\cite{Bijnens:2020xnl} and perturbative
corrections~\cite{Bijnens:2021jqo} to the OPE have recently been
calculated, thereby reducing one source of uncertainty.

There has been much interest in finding a way to satisfy these SDCs
beyond the model solution discussed by MV~\cite{Melnikov:2003xd}. We
proposed a Regge model of pseudoscalar
resonances~\cite{Colangelo:2019lpu,Colangelo:2019uex}, whereas a solution
based on the resummation of a tower of axial-vector resonances in a
holographic model of QCD (hQCD) was put forward in two more recent
papers~\cite{Leutgeb:2019gbz,Cappiello:2019hwh}. A completely different
approach based on a set of interpolants between long and short distance has
been adopted by L\"udtke and Procura (LP)~\cite{Ludtke:2020moa}. Some of these
works either appeared or were published after the WP, where the estimate
about the impact of the SDCs and of the axial-vector contribution is
significantly lower than what was estimated in~\cite{Melnikov:2003xd}.
While none of the most recent papers has criticized the estimate in the WP,
there are statements in~\cite{Cappiello:2019hwh} that their results also agree with
those in~\cite{Melnikov:2003xd}, and with~\cite{Leutgeb:2019gbz}, who in turn conclude that the MV model is not the correct way to implement SDC1, which makes the whole situation
rather confusing. Given the relevance of the SDCs, which currently
represent the largest contribution to the theoretical uncertainty of
$a_\mu^\text{HLbL}$~\cite{Aoyama:2020ynm}, it is important to understand
the differences between these solutions, clarify to what extent they agree and where exactly differences arise, and reassess the current situation.

From a theoretical point of view, the solution based on hQCD is
particularly relevant and appealing as it represents the first hadronic
model of QCD based on axial vectors that exactly satisfies the axial
anomaly and SDC1 in the chiral limit. As we will discuss in
Sect.~\ref{sec:axials} on the basis of general, model-independent
arguments, the solution has to arise from a resummation of an infinite
tower of axial vectors, as it does in the hQCD models and as is expected when fulfilling SDCs with hadronic states~\cite{Peris:1998nj,Bijnens:2003rc}. The model also has the
advantage that the resummation can be performed analytically, but its
simplicity comes at the price of lack of flexibility: once a number of
inputs is used to pin down the free parameters in the model, any further
quantity can be predicted and shows some discrepancies with QCD
phenomenology.  In particular, in the simplest models on which we will focus here, the fulfillment of the asymptotic
constraints in general generates tensions with phenomenological low-energy
constraints~\cite{Leutgeb:2019gbz,Cappiello:2019hwh}.

The solution originally proposed by MV also exactly satisfies the
axial anomaly and SDC1 in the chiral limit, but achieved this goal by a
mere truncation: every hadronic contribution beyond the pion pole for $g-2$
kinematics was simply dropped. Such an approximation is very well justified
for the three-point function $\langle VVA \rangle$, as explicitly shown
in~\cite{Cappiello:2019hwh}. For what concerns the HLbL amplitude the
situation is different: the MV model extrapolates the OPE expression to low
$q_{1,2}^2$, where it cannot be justified. It was first pointed out
in~\cite{Colangelo:2019lpu,Colangelo:2019uex} that the largest contribution
to $a_\mu$ in the MV model comes from the low-energy region, where additional intermediate states would contribute. The first main point of this paper is then  to demonstrate that this conclusion applies to all the recently proposed implementations, summarized in Sect.~\ref{sec:three_approaches}, explaining why there is general consensus on the numerical impact despite significant differences in the implementations themselves, see Sect.~\ref{sec:numerics}.   

For instance, the hQCD and Regge models differ in their use of pseudoscalar vs.\ axial-vector states. In~\cite{Colangelo:2019lpu,Colangelo:2019uex} 
the main motivation for adopting a Regge model of pseudoscalar resonances
is related to a peculiar property of their contribution to the HLbL tensor.
In a dispersive approach~\cite{Colangelo:2017qdm,Colangelo:2017fiz,Hoferichter:2013ama,Colangelo:2014dfa,Colangelo:2014pva,Colangelo:2015ama} the contribution of narrow-width resonances to the
HLbL tensor is in general ambiguous as it depends on the basis in which the
calculation is performed, unless a set of sum rules is satisfied. Only in
the case of pseudoscalars are these sum rules automatically satisfied. The
drawback of using pseudoscalar resonances is that in the chiral limit they
do not couple to the axial current and therefore cannot contribute to the
anomaly. We argued in~\cite{Colangelo:2019lpu,Colangelo:2019uex} that such a
model would nonetheless represent a useful tool to make a realistic
evaluation of the impact of the SDCs in the physical world, i.e., away from
the chiral limit. 

This view has been challenged by MV in~\cite{Melnikov:2019xkq}: they
emphasized the importance of the axial anomaly as an exact constraint in
the chiral limit, and considered its fulfillment essential in order to
make a reliable estimate of the impact of the SDCs on $a_\mu$. By a
detailed comparison between the MV and the hQCD model we will show,
however, that the relevance of the exact axial anomaly in determining the
four-point function is limited to a kinematic region whose impact on the
calculation of $a_\mu^\text{HLbL}$ is very small. This is one of the most
important conclusions of this paper, which extends and confirms the
findings in~\cite{Colangelo:2019lpu,Colangelo:2019uex}, in line with earlier studies of the relevant momentum regions~\cite{Bijnens:2007pz}.

Since each of the models discussed here cannot claim to be a faithful
representation of QCD but at best be a tool to capture the essential
features thereof in connection with a particular aspect of the $a_\mu$ calculation, it is instructive to compare all three of them, even if
they rely on different degrees of freedom to fulfill the SDCs.
Anticipating our conclusions, we will find a satisfactory agreement between
the hQCD and the pseudoscalar Regge models. We will then use the latter to
update our earlier estimate taking into account the recently calculated
perturbative and non-perturbative corrections to the
OPE~\cite{Bijnens:2020xnl,Bijnens:2021jqo}, see Sect.~\ref{sec:OPE_NLO}. This serves only to illustrate the current status,
because we believe that it is possible to incorporate the good theoretical
properties of hQCD models into our dispersive formalism, after developing a
coherent formulation of axial vectors in the narrow-width approximation. This is the direction in which future
work will evolve, as we will sketch in the outlook in Sect.~\ref{sec:outlook}.

\section{The longitudinal OPE and non-renormalization theorems}
\label{sec:OPE}

We concentrate here on the OPE for the
longitudinal amplitude, which concerns only one of the functions in the HLbL
tensor, namely the $\hat \Pi_1$ function introduced
in~\cite{Colangelo:2017fiz}. Only this function contains the contribution
of the pion pole, which can be written as\footnote{We work with
$q_4^2=0$, but $q_4\neq 0$ and, for simplicity, only consider the
isospin-triplet component for now. The Mandelstam variables are defined as $s=(q_1+q_2)^2$, $t=(q_1+q_3)^2$, $u=(q_2+q_3)^2$.}
\begin{align}
\label{Pi1-general}
\hat{\Pi}_1(q_1^2,q_2^2,q_3^2,0;s,t,u) & = \frac{\Fpigg(q_1^2,q_2^2)
  \Fpig(q_3^2)}{s-M_\pi^2}  \notag \\
&+ \tilde{G}(q_1^2,q_2^2,q_3^2,0;s,t,u) \; .
\end{align}
The transition form factor (TFF) of the pion is a single function, which appears both in the
doubly-virtual and in the singly-virtual case:
$\Fpigg(q^2,0)=\Fpig(q^2)$. The function $\tilde{G}$ collects all additional
contributions not containing any poles at $s=M_\pi^2$.

For the muon $g-2$ calculation we need to take the limit $q_4 \to 0$, which
changes the kinematics as follows: $s=q_3^2$, $t=q_2^2$, and $u=q_1^2$, leading to
\begin{align}
\bar{\Pi}_1(q_1^2,q_2^2,q_3^2)&\equiv \hat{\Pi}_1(q_1^2,q_2^2,q_3^2,0;q_3^2,q_2^2,q_1^2)\notag\\
 &=\frac{\Fpigg(q_1^2,q_2^2)
  \Fpig(q_3^2)}{q_3^2-M_\pi^2} \notag\\
  &+ G(q_1^2,q_2^2,q_3^2) \; ,
\label{eq:g-2kin}
\end{align}
where
$G(q_1^2,q_2^2,q_3^2)=\tilde{G}(q_1^2,q_2^2,q_3^2,0;q_3^2,q_2^2,q_1^2)$.
We stress that taking the limit $q_4 \to 0$ starting from the
representation in~\eqref{Pi1-general} unambiguously leads
to~\eqref{eq:g-2kin}. The splitting between the first and second term is
inherited from the splitting between pole term and the rest for general
kinematics, but is nonetheless unique. If one wanted to identify the pole
term directly for $g-2$ kinematics, the first term in~\eqref{eq:g-2kin}
should only have its residue as numerator, thereby separating any
additional $q_3^2$ dependence carried by the TFFs: in other words,
separating the pseudoscalar poles from the vector-meson ones in the TFFs
(which correspond to cuts from $2\pi$, $3\pi$, etc.\ intermediate states,
if one does not take the narrow-width approximation). Both definitions of
the pion pole (for general or $g-2$ kinematics) are possible and the
relation between the two is completely understood. The connection of this
aspect with the SDCs has been discussed in detail in~\cite{Knecht:2020xyr}.
Here we adopt the splitting between pion pole and the function $G$ given
in~\eqref{eq:g-2kin} and concentrate our discussion on the latter.

In the MV limit, $\hat{q}^2\equiv
q_1^2 = q_2^2\gg q_3^2$, $\hat{q}^2\gg \Lambda_\text{QCD}^2$, with no constraints on $q_3^2$,
the function $\bar \Pi_1$ reads: 
\begin{align}
\bar{\Pi}_1(\hat{q}^2,\hat{q}^2,q_3^2) & = \frac{\Fpigg(\hat{q}^2,\hat{q}^2)
  \Fpig(q_3^2)}{q_3^2-M_\pi^2} \notag\\
&+ G(\hat{q}^2,\hat{q}^2,q_3^2)  \; ,  
\end{align}
which can be further simplified taking into account the leading-order OPE for the pion TFF~\cite{Lepage:1979zb,Lepage:1980fj}:
\begin{equation}
\Fpigg(\hat{q}^2,\hat{q}^2)=-\frac{2 F_\pi}{3
  \hat{q}^2}+\Order(\hat{q}^{-3}) \; . \label{asLimit}
\end{equation}
We now carry out the separation between the pion pole in $g-2$ kinematics (in the chiral limit)
from the rest, and rewrite the expression for $\bar \Pi_1$ as follows:
\begin{align}
\bar{\Pi}_1(\hat{q}^2,\hat{q}^2,q_3^2) & =-\frac{2 F_\pi}{3
  \hat{q}^2} \bigg[ \frac{F_{\pi \gamma \gamma}}{q_3^2}
 \notag\\
 &+ \frac{\Fpig(q_3^2)-F_{\pi \gamma \gamma} }{q_3^2} +\Order(M_\pi^2 ) \bigg]
\notag \\
  & + G(\hat{q}^2,\hat{q}^2,q_3^2) +\Order(\hat{q}^{-3})
 \; ,
\end{align}
where
\begin{equation}
F_{\pi \gamma \gamma}:=\Fpig(0)= \frac{1}{4 \pi^2 F_\pi} \; .
\end{equation}
Since we know how the amplitude has to behave in the chiral
limit~\cite{Melnikov:2003xd}:  
\begin{align}
\bar{\Pi}_1(\hat{q}^2,\hat{q}^2,q_3^2)\Big|_{m_q=0}=-\frac{1}{6
  \pi^2} \frac{1}{\hat{q}^2 q_3^2} +\Order(\hat{q}^{-3})\;,
\label{eq:OPEchiral}
\end{align}
we have to conclude that~\cite{Colangelo:2019uex}
\begin{align}
G(\hat{q}^2,\hat{q}^2,q_3^2)\Big|_{m_q=0} &=\frac{2 F_\pi}{3
  \hat{q}^2}\frac{\Fpig(q_3^2)-F_{\pi \gamma \gamma}}{q_3^2}\bigg|_{m_q=0}\notag\\
  &+\Order(\hat{q}^{-3}) \; .
\label{eq:chiral}
\end{align}
This remarkable result is actually a consequence of the non-renormalization of the axial anomaly~\cite{Adler:1969gk,Bell:1969ts,Bardeen:1969md,Wess:1971yu,Witten:1983tw}, as first discussed in~\cite{Melnikov:2003xd} (see~\cite{Vainshtein:2002nv,Knecht:2003xy} for a full account of
non-renormalization theorems for the $VVA$ correlator):
\begin{align}
w_L(q_1^2,q_2^2,(q_1+q_2)^2)=\frac{2 N_C}{(q_1+q_2)^2}+\mathcal{O}(M_\pi^2)
\; .
\label{eq:wLNRT}
\end{align}
The expression on the right-hand side looks like the pion-pole contribution
in the chiral limit, though without the (properly normalized) TFF in the
numerator. As discussed in detail in~\cite{Masjuan:2020jsf,Knecht:2020xyr}
this implies a constraint between the contribution of the pion and that of
transverse degrees of freedom, such that the only effect of their
contribution in the chiral limit is to replace the pion TFF by its value at
$q_3^2=q_4^2=0$.   A crucial point is that the constraint~\eqref{eq:chiral}
applies for arbitrary values of $q_3^2$, but only for large values of $\hat
q^2 \gg \Lambda_\text{QCD}^2$ (and $\hat
q^2 \gg q_3^2$), where the OPE in the MV limit is valid: for
non-asymptotic values of $q_{1,2}^2$ the connection between the four- and
the three-point function gets lost and nothing can be inferred on the
behavior of $G(q_1^2,q_2^2,q_3^2)$.
The key point in assessing 
the relevance of the non-renormalization theorem for
$\langle VVA \rangle$ for the numerical evaluation of $a_\mu$ thus concerns the weight of the integration region in which the constraint applies, as we will discuss in detail in Sect.~\ref{sec:numerics}.

\section{On axial-vector contributions to the longitudinal function in the dispersive approach}
\label{sec:axials}

Since the non-renormalization theorems on the $\langle VVA \rangle$ function
interrelate transverse and longitudinal degrees of freedom, it is clear that
axial-vector states play a role in fulfilling~\eqref{eq:chiral}.
Within the dispersive framework for HLbL, the inclusion of axial-vector mesons
suffers from two closely related difficulties: on the one hand, the contribution of
narrow states depends on the choice of basis---these ambiguities apply to all
narrow resonances beyond the pseudoscalar ones and have been recently
discussed for scalar contributions~\cite{Danilkin:2021icn}.
As only the full HLbL scattering amplitude needs to be basis independent, a
phenomenological evaluation of axial-vector effects thus must proceed in
accordance with the other contributions. The axial-vector exchanges
discussed in the context of SDCs, both in
hQCD~\cite{Leutgeb:2019gbz,Cappiello:2019hwh} and in other
implementations~\cite{Melnikov:2003xd,Melnikov:2019xkq,Roig:2019reh,Masjuan:2020jsf,Pauk:2014rta,jegerlehner:2017gek},
typically refer to a Lagrangian model, which can differ by non-pole pieces
from a dispersive definition, depending on the choice of basis.\footnote{Note that, in addition, all such estimates assume the validity of a narrow-width approximation. Since the main branching fractions proceed into three- or higher-multiplicity final states, a full dispersive treatment of axial-vector intermediate states is difficult, but for $S$- and $D$-wave resonances that decay predominantly into two-meson states a comparison to implementations
in terms of $\gamma^*\gamma^*$ amplitudes~\cite{GarciaMartin:2010cw,Hoferichter:2011wk,Moussallam:2013una,Danilkin:2018qfn,Hoferichter:2019nlq,Danilkin:2019opj,Danilkin:2017lyn,Lu:2020qeo} is possible, see~\cite{Danilkin:2021icn} for the scalar case.} 

The second difficulty in the dispersive approach concerns kinematic singularities:
while the basis of~\cite{Colangelo:2017fiz} is free from kinematic singularities
in the dispersed Mandelstam variable, it still contains singularities
in the photon virtualities, with residues that vanish for the entire HLbL
contribution due to the presence of sum rules. As narrow resonances do not fulfill
the sum rules individually, a further ambiguity in their contribution is introduced
by the subtraction scheme of the singular parts, which again
disappears only for the entire HLbL contribution.
This second complication does not affect pseudoscalar or scalar
contributions, but appears for axial and higher-spin resonances in the basis of~\cite{Colangelo:2017fiz}.
By employing the sum rules, we have now constructed a new basis that explicitly removes all
kinematic singularities from axial-vector contributions, while leaving pseudoscalar
and scalar contributions unaltered, thereby solving this second issue in the case of axial-vector contributions. 
In this basis the contribution of a single axial-vector meson (with mass $M_A$) to the
function $G$ takes the form:
\begin{align}
\label{Gaxial}
G(q_1^2,q_2^2,q_3^2)= \frac{G_2(q_1^2,q_2^2) G_1(q_3^2)}{M_{A}^6}\;,
\end{align}
where
\begin{align}
\label{axialTFFs}
G_2(q_1^2,q_2^2)&=(q_1^2-q_2^2)\F_1(q_1^2,q_2^2)\notag\\
	&+q_1^2\F_2(q_1^2,q_2^2) + q_2^2 \F_2(q_2^2,q_1^2)\;, \nn
G_1(q^2)&=\F_1(q^2,0)+\F_2(q^2,0)=\frac{G_2(q^2,0)}{q^2} \; , 
\end{align}
and $\F_{1,2}(q_1^2,q_2^2)$ are two of the three TFFs of an axial-vector meson, see~\cite{Hoferichter:2020lap} for the precise definitions. The
third one, $\F_3$, does not appear in the expression above but is related to
$\F_2$ by the symmetry properties of the TFFs:
\begin{align}
\F_1(q_1^2,q_2^2)=-\F_1(q_2^2,q_1^2)\; ,\nonumber\\
\F_2(q_2^2,q_1^2)=-\F_3(q_1^2,q_2^2)\;  .
\end{align}
The expression~\eqref{axialTFFs} shows that the dispersive contribution of axial-vector mesons to the function $G$
has the form of non-pole terms, but does not vanish. Our new basis avoids
any kinematic singularities in~\eqref{Gaxial} and makes the dependence on
the virtualities unambiguous for basis changes that preserve this property, up to terms that are subleading for $q_i^2\gg M_A^2$.

As the remaining ambiguities become irrelevant for asymptotic virtualities, \eqref{Gaxial} leads to an interesting model-independent conclusion.
The light-cone expansion determines the asymptotic behavior of $\F_1=\Order(1/q_i^6)$, $\F_2=\Order(1/q_i^4)$, with coefficients determined via decay constants in analogy 
to~\eqref{asLimit}, see~\cite{Hoferichter:2020lap}. This implies that, asymptotically,
\begin{equation}
 G_2(\hat{q}^2,\hat{q}^2)=\Order(1/\hat q^2)\;,\quad 
 G_1(q_3^2)=\Order(1/q_3^4)\;.
\end{equation}

\begin{figure}[t]
 \centering
 \includegraphics[width=\linewidth,clip]{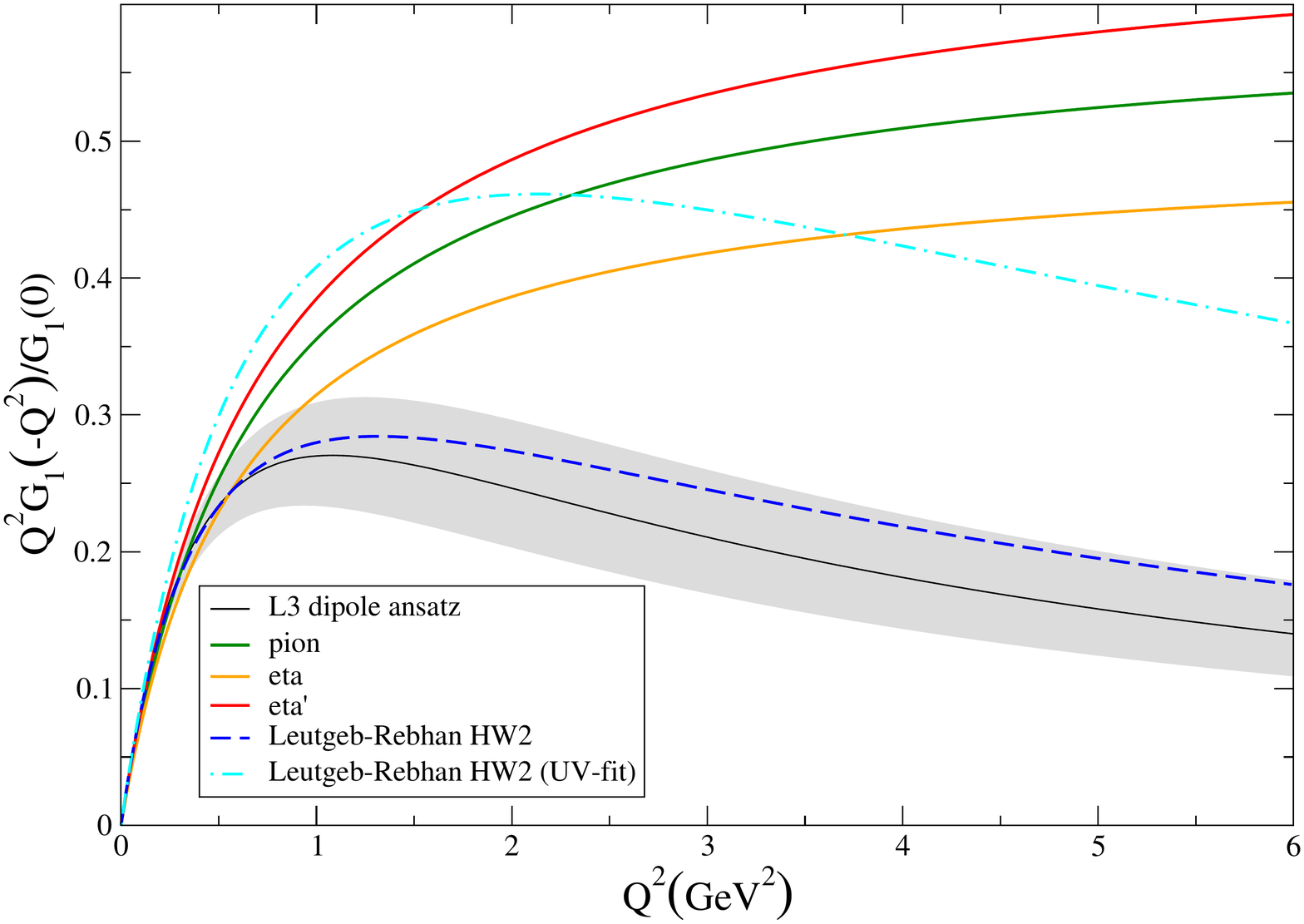}
	\caption{Singly-virtual TFF of the ground-state axial-vector meson $f_1(1285)$:
        comparison of the dipole ansatz used to fit the L3 data (black curve and gray band)~\cite{Achard:2001uu}, the HW2 hQCD
model representations (blue dashed and turquoise dot-dashed curves)~\cite{Leutgeb:2019gbz}, and the
        one obtained from~\eqref{singleA-SDC} using as input the $\pi^0$, $\eta$, and $\eta'$ TFFs (green, yellow, and red curves) from~\cite{Colangelo:2019uex}.}
	\label{fig:axialTFF}
\end{figure}

Since~\eqref{Gaxial} factorizes into parts dependent on $q_{1,2}^2$ and $q_3^2$, respectively, we find that~\eqref{eq:chiral} decomposes into two equations that need to be fulfilled separately:
\begin{align}
\label{singleA-SDC}
\lim_{\hat{q}^2\to \infty}
x \frac{G_2(\hat{q}^2,\hat{q}^2)}{M_{A}^4}& =-\frac{2}{3 \hat{q}^2} +
\Order(\hat{q}^{-3}) \; , \\
\frac{G_1(q_3^2)}{x M_{A}^2}& =-F_\pi \frac{\Fpig(q_3^2)-F_{\pi \gamma \gamma}}{q_3^2} \; ,
\nonumber
\end{align}
with $x$ an unknown, but constant factor that depends on the axial-vector TFFs. While the asymptotic form of $G_2$ thus matches, the axial-vector contribution to $G_1(q_3^2)$ decreases too fast, mirroring the need for an infinite tower of axial-vector  states in the hQCD models. This mismatch can also be illustrated by comparing $Q^2G_1(-Q^2)/G_1(0)$ evaluated from the axial-vector TFFs with~\eqref{singleA-SDC}, see Fig.~\ref{fig:axialTFF}, where we concentrated on the contribution from $\F_2$, given that $\F_1$ is suppressed for several reasons: for small virtualities its antisymmetry implies $\F_1(-Q^2,0)\sim Q^2$, for large virtualities due to the asymptotic behavior, and phenomenologically due to small couplings~\cite{Zanke:2021wiq}. The comparison curves for $\eta$ and $\eta'$ show that this qualitative behavior does not depend on the isospin channel, reinforcing that a single state is not sufficient to implement the SDCs.

\section{Three approaches to satisfy short-distance constraints}
\label{sec:three_approaches}

In this section we compare the different solutions to the SDCs that have
been proposed so far in the
literature~\cite{Melnikov:2003xd,Colangelo:2019lpu,Colangelo:2019uex,Leutgeb:2019gbz,Cappiello:2019hwh,Melnikov:2019xkq},
in terms of the different representations of the functions
$w_L(q^2)\equiv w_L(q^2,0,q^2)$ and $G(q_1^2,q_2^2,q_3^2)$, as constrained by the non-renormalization theorem~\eqref{eq:wLNRT} and the asymptotic
behavior~\eqref{eq:chiral} in the chiral limit. We consider the models as
they are, in other words, we take each one of them as an approximation to
the total contribution to $w_L$ and the longitudinal SDC. The question of
building a better model, possibly by combining features or degrees of
freedom of the present ones will be touched upon in Sect.~\ref{sec:outlook}. The analysis based
on interpolants~\cite{Ludtke:2020moa} will be included in the numerical
comparison in the following section.

\subsection{The Melnikov--Vainshtein model}
After deriving the SDC for the
HLbL tensor, MV go beyond the asymptotic limit and propose a model that by
construction satisfies~\eqref{eq:OPEchiral}:
\begin{align}
\label{eq:MVmodel}
\bar{\Pi}_1^\mathrm{MV}(q_1^2,q_2^2,q_3^2) & = \frac{\Fpigg(q_1^2,q_2^2)
  F_{\pi \gamma \gamma}}{q_3^2-M_\pi^2} \; ,
\end{align}
with the shift in the pole position from zero to $M_\pi^2$ as the only
effect of the light quark masses considered. This implies that,
even though no additional
contributions beyond the pion pole are introduced explicitly, such additional contributions are implicitly assumed to be completely determined not only in the asymptotic region, as
implied by~\eqref{eq:chiral}, but everywhere: 
\begin{align}
G^\text{MV}(q_1^2,q_2^2,q_3^2)= \notag \\
-\Fpigg(q_1^2,q_2^2)&\frac{\Fpig(q_3^2)-F_{\pi
    \gamma \gamma}}{q_3^2} +\mathcal{O}(M_\pi^2) \; . 
\label{eq:MV1}
\end{align}

Equation~\eqref{eq:MV1} is a very strong assumption, with no apparent physical
justification: it extrapolates a constraint only valid at asymptotically
high energies~\eqref{eq:chiral} to all possible values of $q_{1,2}^2$, all
the way down to $q_1^2=q_2^2=0$. As such, it has the potential to
significantly affect the value of the HLbL contribution to $a_\mu$---a
quantity most sensitive to low $q_i^2$. That this indeed happens has
already been shown explicitly in~\cite{Colangelo:2019uex}, and will be
discussed in more detail below.

Regarding the three-point function, the MV model reads
\begin{align}
w_L^\text{MV}(q^2)=\frac{2 N_C}{q^2-M_\pi^2} \; ,
\label{eq:wLMV}
\end{align}
which again amounts to including as only chiral correction the one that shifts
the pole in the pion propagator. Since chiral corrections become negligible
at large $q^2$, \eqref{eq:wLMV} is a well-motivated model at all
$q^2$, with small deviations from the truth expected only at intermediate
energies (chiral corrections may become more sizable for the $\eta$ and $\eta'$ channels).

\subsection{The Leutgeb--Rebhan and Cappiello--Cat\`a--D'Ambrosio--Greynat--Iyer 
   models}\label{HW2sec}
   
 In two recent papers, Leutgeb--Rebhan~\cite{Leutgeb:2019gbz} (LR) and
 Cappiello et al.~\cite{Cappiello:2019hwh} (CCDGI) have proposed models
 based on hQCD to satisfy the SDCs. As the discussion in
 Sect.~\ref{sec:axials} shows, a solution in terms of a single axial-vector
 meson (per isospin channel) is essentially excluded, and indeed in these
 models the solution emerges from a resummation of an infinite tower of
 axial-vector mesons. For simplicity, we concentrate here on the model
 presented by CCDGI in~\cite{Cappiello:2019hwh}, which is equivalent to the
 HW2 model in~\cite{Leutgeb:2019gbz}, although the two groups make
 different choices for the parameters.\footnote{We will not consider the HW1 model
   discussed in~\cite{Leutgeb:2019gbz} simply because it does not offer a
   simple analytic representation like HW2.} The representation of this
 model for the function $G$ reads (in the notation of CCDGI)
\begin{align}
G^\text{HW2}(q_1^2,q_2^2,q_3^2)&=-\frac{\Fpigg(q_1^2,q_2^2)
  \Fpig(q_3^2)}{q_3^2}\\
& - \frac{F_{\pi \gamma \gamma}^2}{q_3^2} \int_0^{z_0}
dz \,\alpha'(z) v_1(z) v_2(z) v_3(z)\;,\notag
\end{align}
where $\alpha(z)=1-z^2/z_0^2$, with $z_0=(\!\sqrt{2} \pi F_\pi)^{-1}$ and 
\begin{equation}
v_i(z)= z Q_i\left[ K_1(z Q_i)+\frac{K_0(z_0Q_i)}{I_0(z_0Q_i)} I_1(z
  Q_i)\right] \; ,
\end{equation}
with $K_n(x)$ and $I_n(x)$ modified Bessel functions~\cite{Cappiello:2019hwh}.
The same function $v_i(z)$ also determines the pion TFF
\begin{equation}
\Fpigg(q_1^2,q_2^2)=-F_{\pi \gamma \gamma} \int_0^{z_0}
dz \,\alpha'(z) v_1(z) v_2(z) \; .
\end{equation}
This representation correctly reproduces the high-$q^2$ limit of the TFF shown in~\eqref{asLimit}, the Brodsky--Lepage limit
of the singly-virtual pion TFF~\cite{Lepage:1979zb,Lepage:1980fj}:
\begin{equation}
\Fpig(\hat{q}^2)=-\frac{2 F_\pi}{
  \hat{q}^2}+\Order(\hat{q}^{-3}), \label{BLlimit}
\end{equation}
as well as, by construction, the normalization at $q_1^2=q_2^2=0$.
A convenient rewriting for $G$ is
\begin{align}
G^\text{HW2}(q_1^2,q_2^2,q_3^2)&=-\Fpigg(q_1^2,q_2^2)
  \frac{\Fpig(q_3^2)-F_{\pi \gamma \gamma}}{q_3^2} \notag \\
&- \frac{F_{\pi \gamma
        \gamma}^2}{q_3^2} \int_0^{z_0} dz \,\alpha'(z) v_1(z) v_2(z)
    \bar{v}_3(z), 
\label{eq:GCCDGI}
\end{align}
where $\bar{v}_3(z)=v_3(z)-1$, as it shows that there is no divergence at
$q_3^2=0$ (the integral vanishes for $q_3^2\to0$). Note that the first term
in~\eqref{eq:GCCDGI} coincides, up to chiral corrections, with the MV
model, which can therefore be viewed as a truncated version of the hQCD
model. In the HW2 model, however, the first and the second non-factorizable
term always come together as they have the same physical origin: both arise from
the resummation of the whole tower of axial-vector mesons. In the numerical
analysis below we will see that, while the first term is dominant for
asymptotic values of $q_1^2\sim q_2^2$, for low momenta, they are
equally important and in fact cancel each other.

We also observe that the HW2 model offers a compact and convenient
representation for the function $\bar{\Pi}_1$:  
\begin{align}
\bar{\Pi}_1^\text{HW2}&= \Fpigg(q_1^2,q_2^2) \frac{F_{\pi
    \gamma\gamma}}{q_3^2-M_\pi^2} \notag\\
    &\times \left[1 + \frac{M_\pi^2(\Fpig(q_3^2)-F_{\pi \gamma
        \gamma})}{q_3^2 F_{\pi \gamma \gamma} }\right] \notag \\
&- \frac{F_{\pi \gamma
        \gamma}^2}{q_3^2} \int_0^{z_0} dz \,\alpha'(z) v_1(z) v_2(z)
    \bar{v}_3(z),
\label{eq:P1HW2}
\end{align}
where one can clearly see that the corrections to the $1/q_3^2$ behavior,
i.e., the pion pole in $g-2$ kinematics, vanish in the chiral limit---the
integral behaves as $\Order(\hat{q}^{-4})$.

Finally, for the $VVA$ correlation function the HW2 model gives
\begin{align}
w_L^\text{HW2}(q^2)=\frac{2 N_C}{q^2-M_\pi^2} \left[1 + \frac{M_\pi^2(\Fpig(q^2)-F_{\pi \gamma
        \gamma})}{q^2 F_{\pi \gamma \gamma} }\right] \; ,
\end{align}
which shows that the first term again corresponds to the MV model. In this
case it is evident that the corrections to the MV model are of
$\mathcal{O}(M_\pi^2)$ for any value of $q^2$, and therefore expected to be
small everywhere. This expression also shows that the first term
in the representation of the four-point function~\eqref{eq:P1HW2} takes the form
$\Fpigg(q_1^2,q_2^2)F_{\pi \gamma \gamma} w_L(q_3^2)/(2N_C)$. While there is
no harm in approximating the $w_L$ function with~\eqref{eq:wLMV}, it is dropping the non-factorizable second term in~\eqref{eq:P1HW2} that amounts to
an uncontrolled approximation. Its numerical impact will be shown in Sect.~\ref{sec:numerics}.

\subsection{Regge model of excited pseudoscalars}
In the model we presented in~\cite{Colangelo:2019lpu,Colangelo:2019uex}, we considered
only the contribution of excited pseudoscalars to the function $G$:
\begin{align}
\label{eq:ourmodel}
G^\text{eP}(q_1^2,q_2^2,q_3^2) &= \sum_{i=1}^\infty
\frac{\FPigg (q_1^2,q_2^2)\FPig(q_3^2)}{q_3^2-M_{P_i}^2}  \; . 
\end{align}
Clearly, by dropping axial-vector intermediate states, which contribute to
this function according to~\eqref{Gaxial}, we are transferring their unique
role in the chiral limit to the pseudoscalars, which amounts to effectively
changing their chiral behavior, in particular the coupling of the
excited pseudoscalars to the axial-vector current, which has to vanish in
the chiral limit. This procedure cannot be strictly justified, but is similar in spirit to
models that use constituent quark masses. In order to remove some of the model
dependence, after matching to the behavior dictated by the OPE we are replacing our model in the asymptotic region
with the perturbative QCD quark loop.

We have imposed as constraint to our model that it satisfy~\eqref{eq:chiral} only for $q_3^2 \gg \Lambda_\text{QCD}$, which is
a less ambitious goal than the one reached by both models
described above:
\begin{align}
\label{eq:ourmodel-asymp}
\lim_{\hat{q}^2 \to \infty } \hat{q}^2 G^\text{eP}(\hat{q}^2,\hat{q}^2,q_3^2) &=
-\frac{1}{6 \pi^2 q_3^2}+\Order(q_3^{-3})\; .
\end{align}
By construction, our model takes into account singularities that are known
to be present in the spectrum of QCD (the low-lying pseudoscalar
excitations). The resummation of all higher excitations is used essentially
to achieve the matching to the asymptotic behavior, but its precise form is
inessential.

The resulting representation for the longitudinal component of the $VVA$ correlator becomes\footnote{Note that here we are using a peculiarity of our model
  that $
F_{P_i
    \gamma^*\gamma^*}(\hat{q}^2,\hat{q}^2)=-\frac{1}{6\pi^2 F_{\pi \gamma \gamma}\hat{q}^2}+\Order(\hat{q}^{-3}) \;$
  for all $i$.} 
\begin{align}
w_L^\text{eP}(q^2)= \frac{2 N_C}{F_{\pi
    \gamma \gamma}}\left[ \frac{\Fpig(q^2)}{q^2-M_\pi^2}+ \sum_{i=1}^\infty
\frac{ \FPig(q^2)}{q^2-M_{P_i}^2} \right] \; .
\end{align}
We stress that this model was not conceived to approximate this function
other than for asymptotic values of its argument, and its use
in~\cite{Colangelo:2019lpu,Colangelo:2019uex} was limited to the four-point
function. However, we find it instructive to provide this expression and compare
it numerically to the other models.

\section{Numerical comparison of the three models}
\label{sec:numerics}
In this section we compare the three models numerically, first for the $w_L(q^2)$ function, then the $G$ function and its
contribution to $a_\mu$. 

\begin{figure}[t!]
\centering
\includegraphics[width=\linewidth,clip]{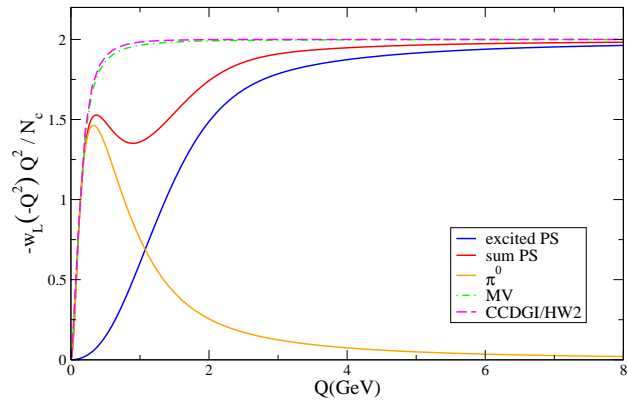}
\caption{\label{fig:wLcomp}
Comparison of the MV, the HW2, and our model
  (solid) for the isovector component of $-Q^2 w_L(-Q^2)$ as a function of
  $Q$. Our model is also broken down into pion and sum of excited
  pseudoscalar contributions.}
\end{figure}

\subsection{The $w_L$ function}
A numerical comparison between the hQCD model and the MV model for the
$w_L$ function was provided and discussed in~\cite{Cappiello:2019hwh}
and clearly showed that the difference between the two models amounts to
chiral corrections neglected by MV, which the hQCD model
estimates to be numerically very small. The picture that emerges from the
comparison is that the $w_L$ function is essentially determined by
its low-energy (fixed by the pion pole) and its high-energy
behavior, with no room for any structure in between. In
Fig.~\ref{fig:wLcomp} we repeat the comparison for the isovector channel and
show in addition the contribution of the pion if one includes its
transition form factor in the numerator---in other words, according to the
dispersive definition of the pion contribution for general kinematics. The
difference between the $\pi^0$ and the CCDGI/HW2 curves is the contribution
of the axials, but its main effect is to remove the TFF from the numerator,
as the minute difference to the MV model shows. The hQCD models confirm that the MV
model appears to be an excellent approximation to the true $w_L$ function
in QCD. It is instructive to see algebraically how the contribution of the
axial-vector states manages to remove the TFF from the pion-pole contribution and
also to be able to estimate the additional corrections to it, but for all
practical purposes, and unless the highest precision is required, the MV
model seems to provide an excellent description of $w_L$.
In the same plot we also show our model of excited pseudoscalars. This is
designed to agree with the other two for asymptotic values of $q^2$, as indeed it does. At low energy, where the pion contribution dominates, it also
agrees with the other two, but the transition region is not as smooth and
shows some structure, with up to 30\% discrepancy with the other two
models. There is nothing to be read into this discrepancy other than the fact
that the model was never designed to provide a good
description of $w_L$: the structure it shows in the intermediate region
just reflects the fact that it was not required to fulfill any
constraints here. If required, the model could be refined to improve the transition between low- and high-energy constraints. Whether this is of relevance in the context of the four-point function will be discussed in the following.

\begin{figure}[t]
 \centering
 \includegraphics[width=\linewidth,clip]{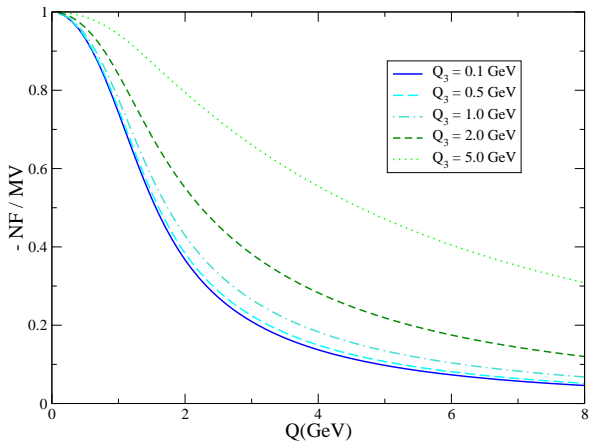}
	\caption{Comparison between the non-factorizable (NF) and the MV term (second and first term in~\eqref{eq:GCCDGI}, respectively, for $q_1^2=q_2^2=-Q^2$ and $q_3^2=-Q_3^2$) in the CCDGI model (set 2).}
	\label{fig:NFvsMV}
\end{figure}

\subsection{The $G$ function and the role of $w_L$}
\label{sec:GandwL}
Such a good understanding of the function $w_L$ and the
simple and accurate description provided by the MV model raises the
question which role the three-point function plays in determining the
four-point function: how strongly does the accurate knowledge of $w_L$
constrain $\bar\Pi_1$ or $G$? MV have shown that for asymptotic values of
$q_1^2\sim q_2^2$, the leading behavior of $G$ is completely fixed by
$w_L$, but how relevant is the asymptotic region in
determining the contribution of $G$ to $a_\mu$? In~\cite{Melnikov:2019xkq} it is argued that the kinematic region $q_1^2\sim
q_2^2 \gg q_3^2$ ``provides the largest contribution to
$a_\mu^\text{HLbL}$,'' but we are not aware of any
quantitative basis for such a statement. 
When we proposed an alternative
way to fulfill the SDCs and compared to
MV~\cite{Colangelo:2019lpu,Colangelo:2019uex}, we showed that the large
difference between ours and the MV model arose precisely in the low-$q_i^2$
region and, moreover, that the largest contribution to $a_\mu^\text{HLbL}$
in the MV model itself came from the same region.

\begin{figure}[t!]
\centering
\includegraphics[width=\linewidth,clip]{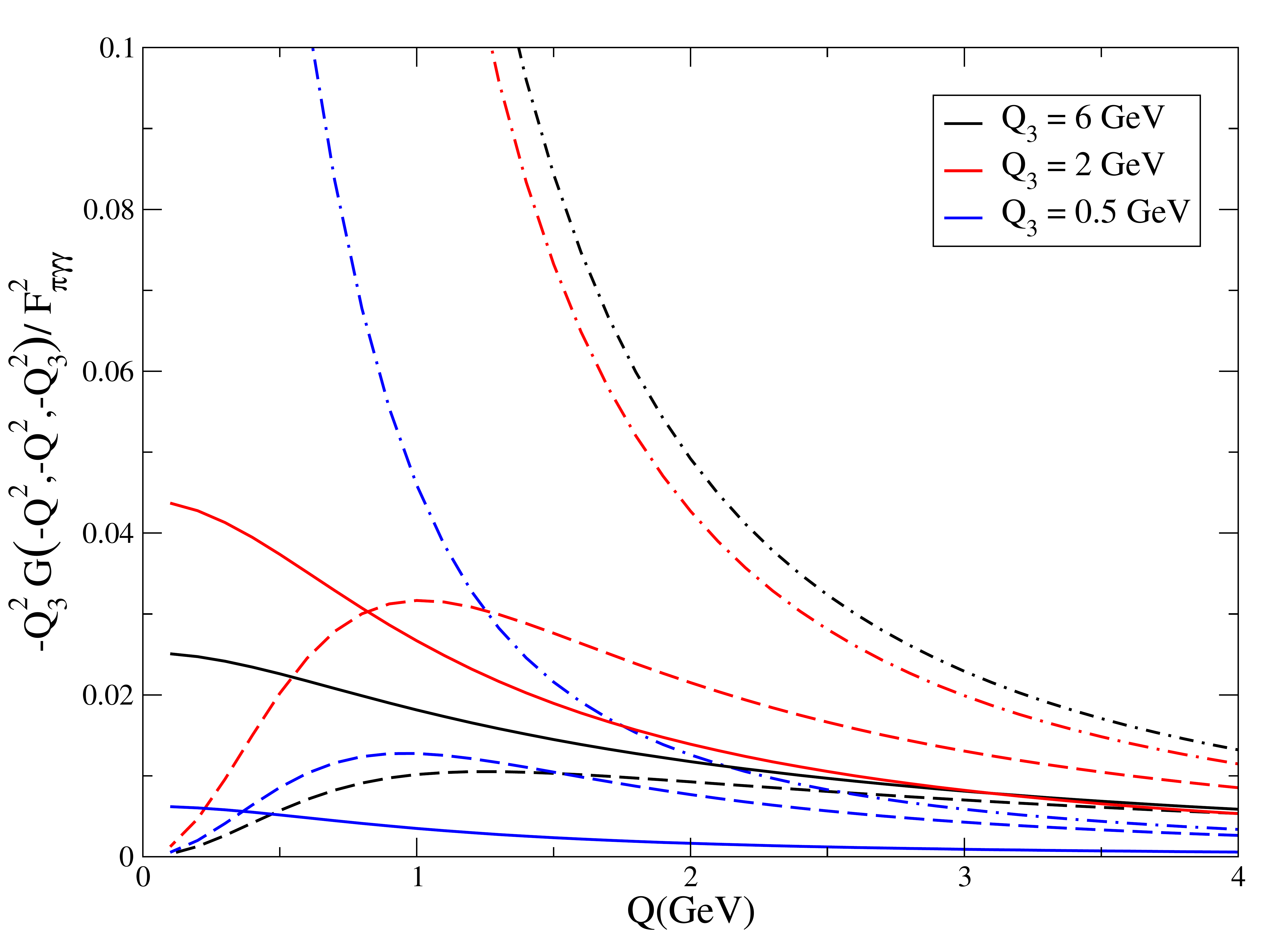}
\caption{\label{fig:Gcomp} Comparison of the MV (dot-dashed), the HW2 (dashed),
  and our model (solid) for $-Q_3^2 G(-Q^2,-Q^2,-Q_3^2)/F_{\pi \gamma
    \gamma}^2$ as a function of $Q$ for different values of $Q_3$.}
\end{figure}

The hQCD models, which satisfy exactly the anomaly and the MV constraints,
allow us to test the approximation made in the MV model in a more
quantitative way. As discussed above, there are two approximations made by MV in
their representation of $\bar\Pi_1$: the first is to neglect chiral
correction in $w_L(q_3^2)$, which is a very good one as we have just seen,
but the second one is to neglect non-factorizable corrections, which in the
hQCD models are given by the integral term in~\eqref{eq:P1HW2}. To
establish the relevance of the function $w_L$ in the calculation of the
four-point function and its contribution to $a_\mu$ we can therefore
compare the non-factorizable and the MV term: the region where the latter
dominates is the region where the MV limit matters, i.e., where the $w_L$
function plays an important role, and the MV model is a good approximation.
This comparison is shown in Fig.~\ref{fig:NFvsMV}: even for the modest
requirement that the non-factorizable term amounts to at most 30\% of
the MV term, the minimum value of $Q$ for which this is satisfied is above
$2\GeV$. At the matching point $Q_\text{match}=1.7$ GeV, adopted
in~\cite{Colangelo:2019lpu,Colangelo:2019uex} for the transition between
the hadronic and the pQCD description, and which will again be used below,
the non-factorizable term is at least a 50\% correction to the MV term.

\begin{table*}[t]
{\small
\begin{center}
\begin{tabular}{lccccccc}
\toprule
& \multirow{2}{*}{MV} &\multicolumn{2}{c}{CCDGI} &
\multicolumn{2}{c}{LR}&\multirow{2}{*}{LP}&\multirow{2}{*}{PS Regge} \\
\cline{3-6}
&&set 1& set 2&HW2&HW2$_\text{UV-fit}$&&\\
\midrule
& \multicolumn{7}{c}{$\Delta a_\mu ^{\pi/a_1} \times 10^{11}$} \\
\midrule
$Q^2_{i}>Q^2_\text{match}\quad \forall i$ & $1.4$& $0.5$ &$0.8$& $0.6$& $0.8$ &$0.9$ & $0.7 $\\ 
$Q^2_{1,2}>Q^2_\text{match}>Q^2_3$        &
$1.4$& $0.8$ &$1.0$  &$0.8$ & $1.0$&$0.3$ &$0.4$\\
$Q^2_{i,3}>Q^2_\text{match}>Q^2_j \quad i \neq j \neq 3$ 
&$0.8$& $0.2$ &$0.3$&$0.2$ & $0.3$ &$0.4$&$0.4$\\
$Q^2_{i}>Q^2_\text{match}>Q^2_{j,k}\quad i \neq j \neq k$&$0.8 $& $0.3$ &$0.4$&$0.3$ & $0.4$ &$0.3$&$0.2$\\
$Q^2_\text{match}>Q^2_{i}\quad \forall i$&$11.8$& $2.2$ &$1.7$& $2.3$ & $1.8$ &$0.7$&$1.0$\\
 \midrule
 Total& $16.2$& $4.0$ &$4.2$& $4.2$ & $4.3$  &$2.6$ &$2.7$\\ 
\midrule
& \multicolumn{7}{c}{$\Delta a_\mu ^{\eta/f_1 + \eta^\prime/f^\prime_1} \times 10^{11}$}\\
\midrule
$Q^2_{i}>Q^2_\text{match}\quad \forall i$					&	$3.4$	& 	$1.4$	& 	$1.7$	& 	$1.7$	& 	$2.5$	& 	$2.5$ &$3.1$\\
$Q^2_{1,2}>Q^2_\text{match}>Q^2_3$       				&	$2.1$	& 	$2.1$	& 	$2.3$	& $2.5$	& 	$3.0$	& 	$0.6$ &$1.1$\\
$Q^2_{i,3}>Q^2_\text{match}>Q^2_j \quad i \neq j \neq 3$		
&	$1.9$	& 	$0.6$	& 	$0.7$	& 	$0.6$	& 	$0.9$	& 	$1.2$ &$1.6$\\
$Q^2_{i}>Q^2_\text{match}>Q^2_{j,k}\quad i \neq j \neq k$		&	$1.7$	& 	$0.8$	& 	$0.9$	& 	$0.9$	& 	$1.1$	& 	$0.7$ &$0.9$\\
$Q^2_\text{match}>Q^2_{i}\quad \forall i$					&	$12.9$	& 	$5.6$	& 	$5.1$	& 	$6.8$	& 	$5.4$	& 	$1.5$ &$3.1$\\
 \midrule
 Total& $22.1$& $10.4$ &$10.7$&$12.5$& $12.8$ & $6.5$ &$9.9$\\ 
 \midrule
 Grand total ($\pi/a_1+\eta/f_1+\eta^\prime/f^\prime_1$)& $38.3$& $14.3$ &$14.9$&$16.7$& $17.1$&$9.1$ & $12.6$ \\
 \bottomrule
\end{tabular}
\end{center}
\caption{\label{tab:numerics} Contribution of $G$ to $a_\mu$ (referred to as the longitudinal SD contribution in~\cite{Colangelo:2019lpu,Colangelo:2019uex} and the longitudinal axial-vector contribution in~\cite{Leutgeb:2019gbz,Cappiello:2019hwh}) from the isovector and isoscalar plus isosinglet channels broken down in
  different integration regions ($Q_\text{match}=1.7\GeV$). 
  The notation for the mixed regions includes the respective crossed versions, e.g., the second line gives the contribution from $\bar\Pi_1$ in the region $Q_{1,2}^2 > Q_\text{match}^2 > Q_3^2$ and from $\bar\Pi_2$ in the region $Q_{1,3}^2 > Q_\text{match}^2 > Q_2^2$, in such a way that the region in which the SDC1 applies is contained in this (and partly the first) row, while the third row has a scaling $1/Q^4$ in the hard momenta. 
  Due to different mixing patterns the $\eta/f_1$ and $\eta^\prime/f_1^\prime$ contributions cannot be compared separately. Note that the Regge-model contribution to the asymptotic region is not yet replaced by the OPE result. The numbers for LP refer to the ``reference interpolant'' of~\cite{Ludtke:2020moa}. The HW1 model, which we
  have not considered here, gives a higher contribution $\Delta a_\mu=23.2\times 10^{-11}$~\cite{Leutgeb:2019gbz}. All entries are understood to be accurate at the level of $\pm 0.1$ due to the applied numerical integration methods, other (model-dependent) errors are not shown. }
}
\end{table*}

This suggests that if we compare the MV and the hQCD models at the level of
the function $G$, the agreement is going to be much worse than for the
function $w_L$. To verify this expectation, we plot the isovector component of the
function $-Q_3^2 G(-Q^2,-Q^2,-Q_3^2)$ as a function of $Q^2$, see
Fig.~\ref{fig:Gcomp}.  The dashed curves show the HW2 model, whereas the
dot-dashed ones only show the first term of~\eqref{eq:GCCDGI}, which corresponds
to the MV model. The plot shows very clearly that the latter two versions
tend to the same asymptotic limit, even for low values of $Q_3$, but that
they differ significantly at low $Q^2$. While at large $Q^2$ the second term
in~\eqref{eq:GCCDGI} is subdominant and can be neglected, it compensates
exactly the first one for low $Q^2$, so that their sum vanishes. This is
expected for axial-vector mesons, and is already seen in
Fig.~\ref{fig:Gcomp}. If one keeps just the first term, i.e., the MV
model, this grows at low $Q^2$, reaching a finite limit for $Q^2=0$ (not
visible in the plot range, because it is quite large).
Fig.~\ref{fig:Gcomp} also displays our model (solid curves), which has
non-vanishing limits for $Q^2=0$, too.  This is well understood, however,
because excited pseudoscalars do couple to two real photons (even in the
chiral limit). As detailed in~\cite{Colangelo:2019uex}, these couplings are
compatible with the available phenomenological information, which still
suffers from large uncertainties. By construction, our model also agrees
with the other two for large $Q^2$ and $Q_3^2$, whereas for low $Q_3^2$ it
does not agree well with the other two even as $Q^2$ grows.

\subsection{Contribution to $a_\mu$ of the function $G$}

We can now address the question of how these differences are reflected in
the calculation of the contribution to $a_\mu$. We do so by breaking down
the contributions from different kinematic regions and separating the
isovector channel from the isoscalar and isosinglet ones. Identifying
the isoscalar and isosinglet pieces with the physical states ignores mixing
effects, which implies that the $\eta/f_1$ and $\eta^\prime/f_1^\prime$
cannot be compared separately. In general, the correct implementation of
mixing effects requires two mixing angles (see~\cite{Gan:2020aco} for a
review), but the differences can be illustrated based on the simple U(3)
formula
\begin{equation}
\frac{\Gamma(P\to \gamma\gamma)}{\Gamma(P'\to\gamma\gamma)}=\frac{M_P}{M_{P'}}\cot^2\big(\theta_A-\theta_0),\, \theta_0=\arcsin\frac{1}{3},    
\end{equation}
which for $P=f_1$ gives
$\theta_A=62(5)^\circ$~\cite{Achard:2001uu,Achard:2007hm}, but
$\theta_A=84.8(6)^\circ$ for $P=\eta$. For this reason, we only compare the
sum of $\eta$ and $\eta'$ with the sum of $f_1$ and $f_1'$ contributions,
which are not affected by this ambiguity.  Table~\ref{tab:numerics} shows
that, although the CCGDI/HW2 and our model completely differ in the degrees
of freedom that are used to satisfy the relevant SDCs, they give similar
numerical contributions to $a_\mu$. The MV model, which
satisfies~\eqref{eq:chiral} exactly, much like CCGDI/HW2, but by neglecting
any contribution to $\bar\Pi_1$ beyond the pion pole in $g-2$ kinematics,
gives instead a much larger contribution.

The breakdown of the contribution to $a_\mu$ in different integration
regions shows that there is in general a rather good agreement (with a few
exceptions) between the Regge and the CCDGI/HW2 models. In particular in
the pion/$a_1$ channel, the agreement is very good in the ``asymptotic''
region, the first row in the table. At low $q^2$ there are differences, but
these are expected, because the two models describe different degrees of
freedom there. The situation is similar in the $\eta/f_1 + \eta^{\prime
}/f_1^\prime$ channels, where again the largest differences occur in the
low-energy region. However, there are also some non-negligible differences
even in the large-$Q_i^2$ region, which might be related to the fact that
the HW2 models do not fully saturate the SDC2~\cite{Cappiello:2019hwh}. Overall, the PS Regge and also the hQCD models are largely compatible with the LP interpolants, which are independent of the choice of degrees of freedom. 
As far as the MV model is concerned, all regions where at least one of the
$Q_i$ is large are in reasonable agreement with the other two models,
but it is the region where all $Q_i$ are small where the MV
model estimates significantly larger effects; as expected, since in this region the truncation of the non-factorizable contributions, see Sect.~\ref{sec:GandwL}, cannot be justified. The
table also shows that the kinematic region $Q_1^2\sim Q_2^2 \gg Q_3^2$,
$Q_{1,2}^2 > Q_\text{match}$, all contained in part of the first row and in
the second row, provides a small contribution to the total. This is
particularly true for the MV model.

\begin{figure}[t]
 \centering
 \includegraphics[width=\linewidth,clip]{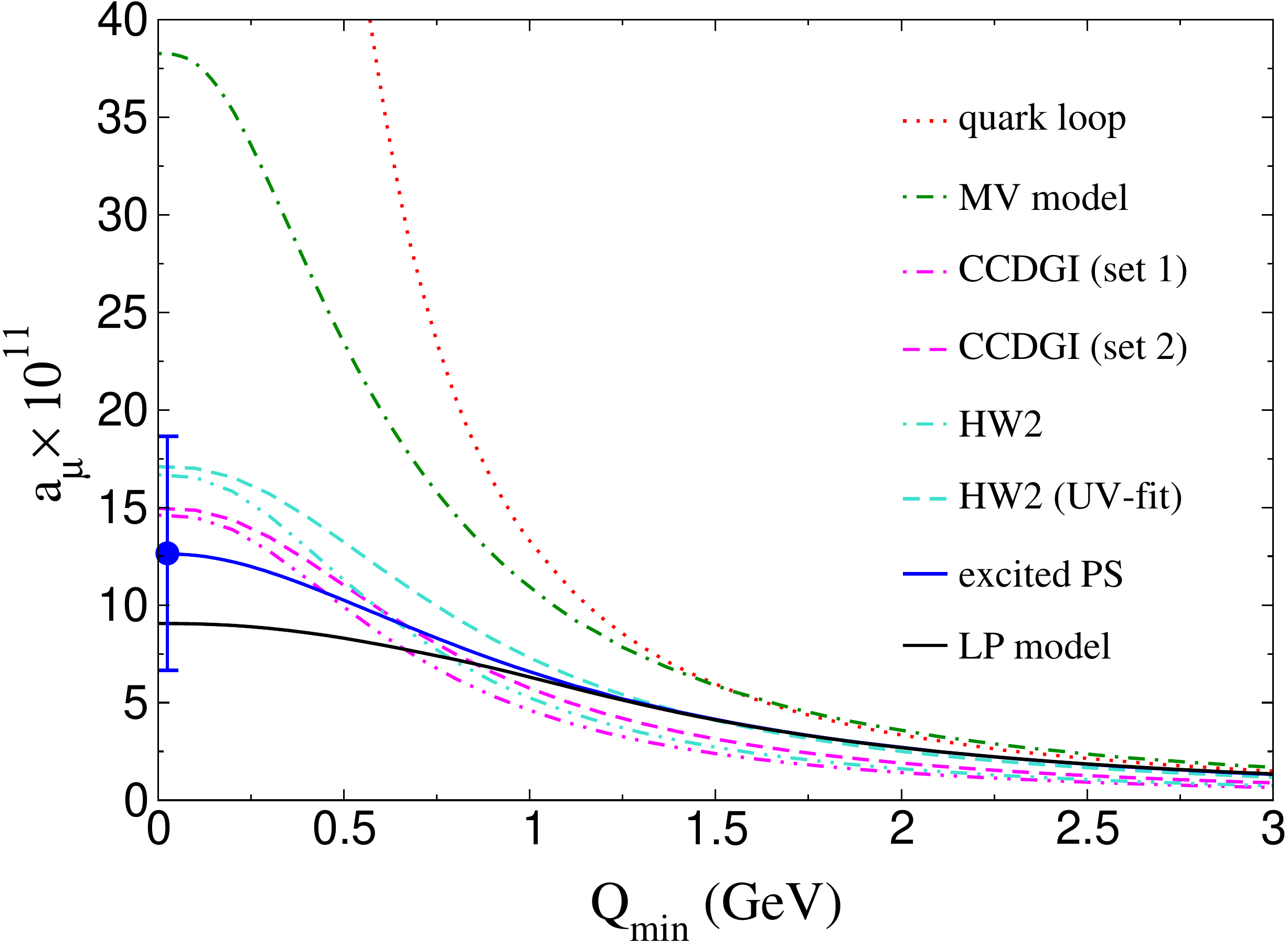}
	\caption{Contribution to $a_\mu$ for $Q_i\geq Q_\text{match}$: the longitudinal part of the massless perturbative QCD quark loop (dotted red), the MV model (dot-dashed green), the CCDGI model (dashed and dotted-dashed magenta), the LR HW2 model (dashed and dotted-dashed turquoise), the LP model (solid black), and our model (solid blue). The blue point indicates the final value in~\eqref{final_result}.}
	\label{fig:amuQmin}
\end{figure}

Another way to visualize the impact on $a_\mu$ of the different kinematic
regions is to plot the contribution to $a_\mu$ as a function of a lower
cutoff $Q_\text{min}<Q_i$, as shown in Fig.~\ref{fig:amuQmin}. The plot shows
again that although there are differences in the $Q_i$ dependence between
the Regge and the hQCD models, these are not so significant with respect to
$a_\mu$ and lead to a similar final number. The MV model, on the other
hand, only comes close to the other two for large values of $Q_\text{min}$,
whereas it estimates a much larger effect in the region of low $Q_i$.
Finally, we have also shown the result obtained with the interpolants by
LP, which is compatible with both the Regge as well as
the hQCD models, even though somewhat lower for low-$Q^2$.  This may have to do with the
fact that it does not include any explicit resonances and lacks the
corresponding low-$Q^2$ enhancements.  However, as explained
in~\cite{Ludtke:2020moa}, the method of interpolants can be generalized to
explicitly include resonance contributions, once their model-independent
description becomes available (and might offer a valuable alternative to
the resummation of a tower of states). For axial-vector states this is not yet the
case, however: a phenomenologically driven evaluation seems within reach at
least for the $f_1$ contribution~\cite{Zanke:2021wiq}, but it will require
a detailed understanding of sum-rule ambiguities.

The present numerical comparison seems to be at odds with the conclusions
drawn by CCDGI~\cite{Cappiello:2019hwh}, who claim to be in agreement with
the MV estimate. They reach this conclusion on the basis of two
comparisons: a detailed one at the level of the $\langle VVA \rangle$
correlation function and one at the level of the total contribution to
$a_\mu$. The first one has been discussed above and indeed shows that the
two models agree very well. However, the
comparison of the contribution to $a_\mu$
at the level of the total without separation of the poles of the ground-state pseudoscalars risks to be misleading: the total number
is dominated by the poles due to the Goldstone bosons and even small
differences in the evaluation of the latter (necessary because of our
improved understanding of their TFF) may obscure the comparison for the
remainder. LR~\cite{Leutgeb:2019gbz}, whose model coincides algebraically
with that of CCDGI and numerically differs very little, make the comparison
after first subtracting the Goldstone-boson poles and come to the same
conclusion we reached here.

\section{Impact of the perturbative corrections to the OPE}
\label{sec:OPE_NLO}

\begin{figure}[t]
 \centering
 \includegraphics[width=\linewidth,clip]{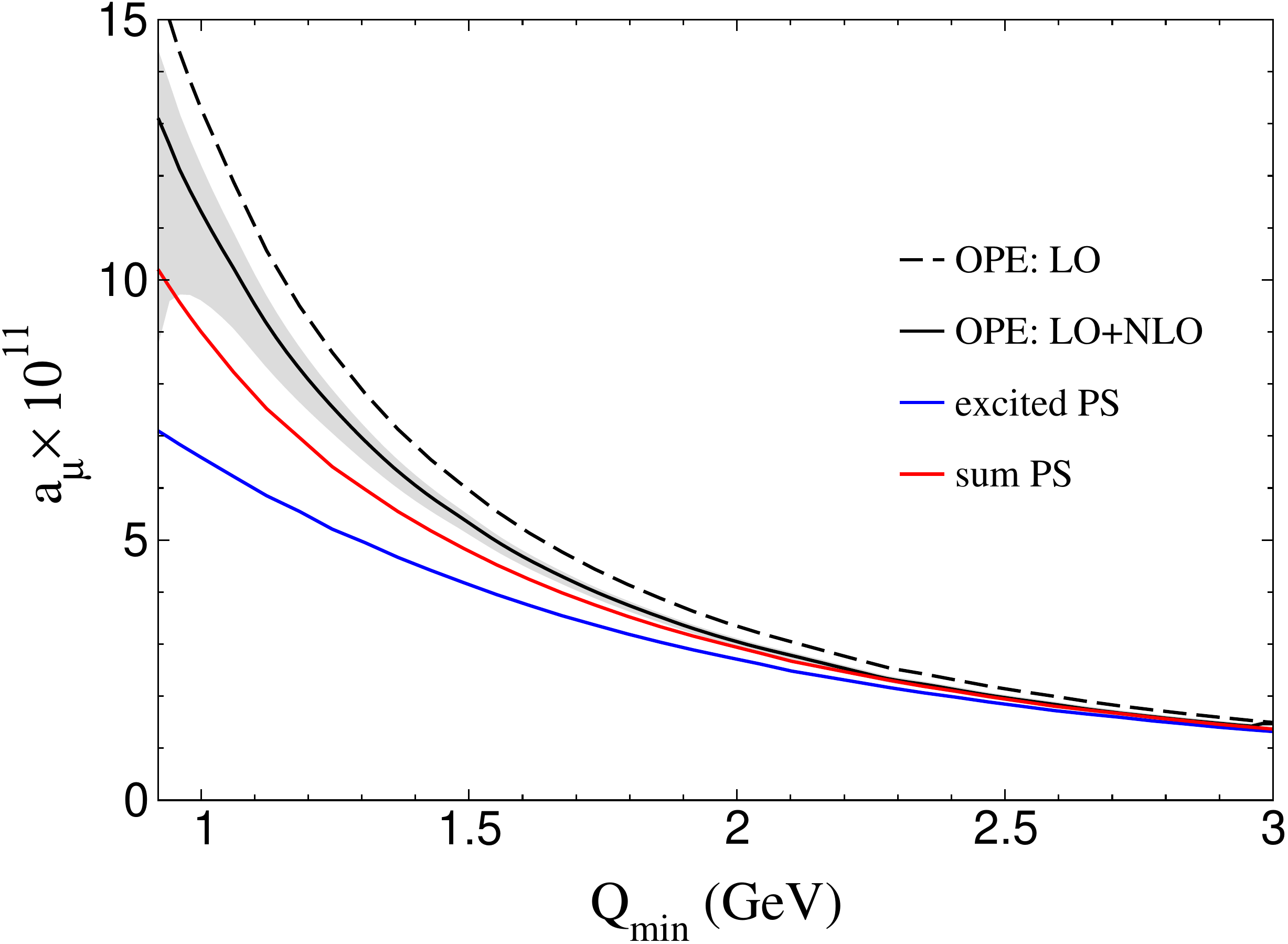}
	\caption{Matching between the  NLO OPE and the Regge model for
          pseudoscalars (red curve). The blue curve shows the contribution
          of only the excited pseudoscalars, excluding the $\pi^0$, $\eta$,
          and $\eta'$. The gray band refers to the uncertainty in setting the $\alpha_s$ input, see~\cite{Bijnens:2021jqo} for more details.}
	\label{fig:amuQminGluons}
\end{figure}

We can now evaluate the effect of the recently calculated gluonic
corrections to the OPE~\cite{Bijnens:2021jqo}, as illustrated in
Fig.~\ref{fig:amuQminGluons}. The NLO corrections lead to a reduction of
the massless quark loop that for integrated quantities tends to evaluate
around $1-\alpha_s/\pi$, e.g., one has at the symmetric point
$Q_1=Q_2=Q_3=Q$~\cite{Bijnens:2021jqo}
\begin{align}
\frac{\bar\Pi_1\big|_\text{NLO}}{\bar\Pi_1\big|_\text{LO}}&=C_1 \frac{\alpha_s}{\pi}\;,\notag\\
C_1&=-\frac{75\Delta^{(1)}-2\Delta^{(3)}-360\zeta_3}{54}\approx - 0.86\;,
\end{align}
where $\Delta^{(n)}=\psi^{(n)}(1/3)-\psi^{(n)}(2/3)$ in terms of the
polygamma function $\psi^{(n)}$ and $\zeta_3\approx 1.202$ (the coefficient of
the $\alpha_s$ corrections becomes exactly $-1$ in the MV
limit~\cite{Ludtke:2020moa}). In the following, we use the full corrections
from~\cite{Bijnens:2021jqo}.

As the hQCD models and the Regge model agree reasonably well at the numerical level, we will rely only on the latter in the following. We
essentially repeat the matching procedure described
in~\cite{Colangelo:2019uex}, replacing the plain massless quark loop with
the one containing $\Order(\alpha_s)$ corrections. The resulting shift,
even down to momentum cutoffs of about $1\GeV$ is small, as can be seen in
Fig.~\ref{fig:amuQminGluons}, but the most relevant improvement is the
reduction of the uncertainties, which had been estimated to be
$\Order(20\%)$ of the massless quark loop in~\cite{Colangelo:2019uex} and
is reduced to a few percent after the NLO
calculation~\cite{Bijnens:2021jqo}, much smaller than the uncertainties on
the hadronic side. As a consequence, the procedure to determine the
matching point by minimizing the total uncertainties would not work
anymore. Instead, we keep it fixed at $1.7(5)\GeV$ as we did
in~\cite{Colangelo:2019uex}. With these changes, our updated estimate of
the impact of longitudinal SDCs on $a_\mu$ reads
\begin{align}
 \Delta
 a_\mu^\text{LSDC}&=\left[8.7(5.3)_{\text{PS-poles}}+4.2(1)_{q\text{-loop}}\right]\times 10^{-11} \notag \\
&= 13(5)\times 10^{-11}\;,
\label{final_result}
\end{align}
where the first number in brackets is the contribution from the region
below the matching momentum of $1.7\GeV$, evaluated as resummation of excited
pseudoscalars, and the second from the region above $1.7\GeV$, evaluated with
the NLO quark loop. A welcome feature of the perturbative corrections is
that they push the OPE curve down, thereby improving the matching with the
hadronic model, which now seems to work perfectly already around $2\GeV$. 

If one takes into account the large uncertainties on the hadronic curve
(not shown in Fig.~\ref{fig:amuQminGluons}) and the rather small estimated perturbative and
non-perturbative uncertainties, one would be led to push the matching point
towards $1\GeV$: this would reduce the importance of both hadronic and
model uncertainties and lead to smaller total uncertainties. To illustrate
the point we mention the number we get for a matching point at the lower
end of the range we considered, for $1.2\GeV$: $ \Delta 
 a_\mu^\text{LSDC}=\left[5.7(2.8)_{\text{PS-poles, par}}+8.1(5)_{q\text{-loop}}\right]\times 10^{-11}
= 14(3)\times 10^{-11}$,
where the error in the hadronic model only refers to the parametric uncertainty---reduced from the $3.6\times 10^{-11}$ it contributes to~\eqref{final_result}---while the remainder of the error estimate, especially the variation of the matching point, does not adapt in a straightforward way to the lower scale. Most notably, the central value only changes slightly, well within the uncertainties of the hadronic model, which shows that the information coming from the perturbative side agrees with the hadronic estimate. Of course, it is
not obvious that at such low energies power corrections beyond the ones
calculated in~\cite{Bijnens:2020xnl} remain irrelevant, and to be on the
safe side one may consider increasing a bit the uncertainties on the OPE
side (also perturbative corrections, estimated in~\cite{Bijnens:2021jqo} via scale variation in $\alpha_s$, could play a role at such low energies). In any case, this brief discussion is just meant to underline the
relevance of the calculation of the corrections to the massless quark loop
and their possible impact in reducing the uncertainties of this
contribution: a full implementation of this is left for future work.

\section{Conclusions and outlook}
\label{sec:outlook}

In this paper we have discussed our current understanding of the role and
impact of longitudinal SDCs on the HLbL contribution to $(g-2)_\mu$ and
updated it to take into account the recent calculation of the NLO
perturbative corrections to the OPE~\cite{Bijnens:2021jqo}. On the
low-energy, hadronic side different solutions for the matching to the SDCs
have been proposed, sometimes accompanied by contradicting statements. To
clarify the situation we have compared these models both at the level of
the longitudinal component $w_L$ of three-point function $\langle VVA
\rangle$ and in terms of the function $G$, defined
in~\eqref{eq:g-2kin}, which collects all contributions beyond the pion pole
to the $\bar\Pi_1$ function of the HLbL tensor---the only one relevant for
the longitudinal SDCs. In this way, the core assumptions and features of
each implementation become most transparent, facilitating the comparison of
the different proposed solutions.

Our conclusions can be summarized as follows:
\begin{enumerate}
\item
  Both the original MV model and the recent hQCD models satisfy the
  axial anomaly in the chiral limit exactly. When compared at the level of
  the three-point function and the longitudinal component $w_L(q^2)$, they
  agree very well, supporting that the MV model is an excellent
  approximation to QCD for this particular quantity. We have also compared
  our Regge model for excited pseudoscalars~\cite{Colangelo:2019uex} and
  showed that, as expected, it satisfies the axial anomaly only
  asymptotically and at low $q^2$.
\item A comparison between the MV and the hQCD models for the four-point
  function, and in particular the function $G$, can be done analytically
  and is very transparent: the MV model can be viewed as a truncation of
  the hQCD models and amounts to dropping all contributions beyond the pion
  pole for $g-2$ kinematics. In the hQCD models these additional
  contributions are expressed in terms of a single integral over Bessel
  functions, which cannot be factorized into a function of $q_{1,2}^2$ and
  one of $q_3^2$. We have analyzed the relative importance of the
  non-factorizable and the MV term and shown that the latter is dominant
  only for rather large values of $q_{1,2}^2$: neglecting the former term
  leads to a significant overestimate of the low-$q_{1,2}^2$ contribution.
  The hQCD and the MV model, which agree almost exactly on the axial
  anomaly, thus differ substantially in their estimate of $a_\mu^\text{HLbL}$.
\item The two approaches that achieve a matching to the OPE by resumming a
  tower of hadronic states provide very similar estimates of the impact on
  $a_\mu$, even though one is based on excited pseudoscalars in a Regge
  model~\cite{Colangelo:2019uex} and the other on axial-vector mesons in a
  hQCD model~\cite{Leutgeb:2019gbz,Cappiello:2019hwh}, with the aforementioned differences in $w_L(q^2)$ in the transition region between low and high momenta. This again shows
  that the role of the axial anomaly in determining the HLbL amplitude and
  its contribution to $a_\mu$ is rather limited.
\item 
The hQCD models~\cite{Leutgeb:2019gbz,Cappiello:2019hwh}
  provide an explicit, analytic solution of the SDC1 in terms of a tower of
  axial-vector resonances, which offers useful insights in the mechanism by which SDC1 is fulfilled. 
Simple versions of these models, such as HW2, depend on very few parameters, which can be pinned down by imposing a number of phenomenological constraints, but once this is done further comparisons to phenomenology show discrepancies. This can be improved by considering more complicated versions of these models, such as HW1.
\item In the chiral limit the axial-vector mesons have to play an important
  role in satisfying the SDCs, and the hQCD models provide a
  concrete realization of the underlying mechanism. In the future it will
  be critical to achieve a full, model-independent understanding of how
  axial-vector resonances contribute to HLbL (in analogy to scalar
  states~\cite{Danilkin:2021icn}), at least in the narrow-width
  approximation. Otherwise a combination with other contributions
  to HLbL scattering evaluated within a dispersive approach would not be
  justified. Here, we have presented the expression for the dispersive
  axial-vector contribution to $\bar\Pi_1$ in a particular choice of basis that is
  compatible with all contributions evaluated dispersively so far, but
  sum-rule ambiguities that are reflected in a basis dependence still need
  to be addressed together with a numerical analysis based on any TFF
  input.
\item The final estimates of $\Delta a_\mu^\text{LSDC}$ obtained with
  the hQCD and our Regge model agree quite well with each other as well
  as with a solution of the SDCs based on
  interpolants~\cite{Ludtke:2020moa}. On this basis, we have updated the
  final result given in~\cite{Colangelo:2019uex} to incorporate the
  perturbative corrections to the OPE calculated in~\cite{Bijnens:2021jqo}:
\begin{equation}
\Delta a_\mu^\text{LSDC}= 13(5)\times 10^{-11} \, .
\end{equation}
Even with the slight reduction of the total uncertainty, this covers all
realistic estimates of the impact of the longitudinal SDCs on
$a_\mu^\text{HLbL}$ present in the literature.
\end{enumerate}

As we argued above, further reductions of the uncertainties in the HLbL
contribution due to the fulfillment of the SDCs are possible, also in view
of the smallness of the perturbative corrections to the OPE and their
uncertainties down to $\sim 1\GeV$~\cite{Bijnens:2021jqo}. This will
require an improved and less model-dependent description on the hadronic
side before trying to optimize the matching and exploiting at best the
result of the perturbative calculation. Some of the recent developments
discussed here have paved the way to this goal. Future steps in this
direction include: 
\begin{itemize}
\item[{\it i)}] fully clarifying how to evaluate the
contribution of axial-vector resonances to $a_\mu^\text{HLbL}$ in an
unambiguous way; 
\item[{\it ii)}] understanding how to incorporate the solution of
the SDCs in the chiral limit provided by the hQCD models in a more general,
dispersively motivated framework based on axial-vector mesons; 
\item[{\it iii)}] while
our discussion here was only concerned with the SDC for the longitudinal
amplitude, a solution in terms of axial vectors can address at the same
time both the longitudinal and the transverse SDCs;
\item[{\it iv)}] once the treatment of axial-vector mesons in the
general dispersive formalism will become possible, the reasons to use a
Regge model of pseudoscalars as a tool to estimate the impact of the SDCs
will cease to exist: only the few lightest excited pseudoscalars will need
to be included. 
\end{itemize}
Work along these lines is ongoing.

\begin{acknowledgements}
We thank Jan L\"udtke and Massimiliano Procura for discussions and for providing numbers about their model. We further thank Antonio Rodr\'iguez-S\'anchez for help with the
numerical evaluation of the perturbative correction to the OPE, and Luigi Cappiello, Oscar Cat\`a, Giancarlo D'Ambrosio, Nils Hermansson-Truedsson, Abhishek Iyer, Marc Knecht, Josef Leutgeb,
Jan L\"udtke, Massimiliano Procura, and Anton Rebhan for comments on the manuscript.  
Financial support by 
the SNSF (Project Nos.\ PCEFP2\_181117 and PZ00P2\_193383 and Grant No.\
200020\_175791) is gratefully acknowledged.
\end{acknowledgements} 

\end{sloppypar}

\bibliographystyle{utphysmod.bst}
\balance
\bibliography{Literature}

\end{document}